# Dynamics of a hybrid cubic vibro-impact oscillator and nonlinear energy sink

Maor Farid

*Massachusetts Institute of Technology, 77 Massachusetts Ave., Cambridge, MA 02139, United States*
*Faculty of Mechanical Engineering, Technion – Israel Institute of Technology, Haifa 3200003, Israel*
*faridm@mit.edu*

**Abstract**

Various dynamical engineering systems involve purely nonlinear stiffness that lacks linear properties even under the assumption of small displacements. The most common is the cubic nonlinearity that may stem from either geometric or material properties. Examples include flexible structural components, pre-tensioned cables, springs, and polymers. When subjected to tight rigid constraints collisions might take place, leading to an additional infinitely strong nonlinearity. The resulting dynamical regime involves both smooth nonlinear oscillations (SNOs) and a hybrid cubic vibro-impact (HCVI) regime. The adaptive nonlinearity of the HCVI oscillator is used by the HCVI-nonlinear energy sink (NES) as an effective mechanism for efficient vibration mitigation in broad energy and frequency ranges. Due to the multiple essential nonlinearities of the HCVI oscillator, traditional analytical methods are inapplicable for the description of its transient dynamics. In the current work, we model the HCVI oscillator by a forced particle in a hybrid quartic-square potential well with infinite depth under periodic external forcing. The slow-flow dynamics of the system is analyzed in the framework of isolated resonance approximation by canonical transformation to action-angle (AA) variables and the corresponding reduced resonance manifold (RM). Two types of bifurcation are examined and described analytically. The former is associated with the SNO-regime and the HCVI-regime and vice versa, and the latter with reaching a chosen maximal transient energy level. Unlike previous studies, three underlying dynamical mechanisms that govern the occurrence of bifurcations are identified: *two* maximum mechanisms and one saddle mechanism. The first two correspond to a gradual increase in the system's response amplitude for a proportional increase in the excitation intensity, and the latter corresponds to an abrupt increase in the system's response and therefore more potentially hazardous when takes place in engineering systems and more effective for energy pumpingwhen takes place in the HCVI-NES. Both mechanism types are universal for systems that undergo escape from a potential well. The maximal transient energy is predicted analytically over the space of excitation parameters and described using iso-energy contours. The response curves of the system are obtained analytically, allowing a full perspective on the HCVI-oscillator dynamical response, and the HCVI-NES performances and design optimization. All theoretical results are supported by numerical simulations.

*Keywords:* Hybrid nonlinear energy sinks, Vibro-impact dynamics, Equivalent potential wells, Action-angle variables,

## 1. Introduction

When various engineering systems are exposed to external disturbances, they might develop strongly nonlinear behavior that involves collisions, impacts, and chattering. Examples include machine and structural components subjected to kinematic constraints. Some of those systems are fundamentally nonlinear and lack linear stiffness even under the assumption of small oscillations. In most cases and due to symmetricity, the stiffness is associated with a cubic term, which can stem from either geometric or material properties. For example, a mass mounted on pre-tensioned wires [1] and conical springs [2], a configuration of multiple linear springs [2], and more. In this case, smooth nonlinear oscillations (SNOs) take place, when the frequency of oscillations is amplitude-dependant. When those systems are subjected to external disturbance and rigid geometric constraints, due to rigid barriers, for example, a hybrid dynamical regime that involves both SNOs and sustained collisions might take place. Hence, this regime is referred to as the hybrid cubic vibro-impact (HCVI) regime. In most cases, this intensive regime can lead to substantial accelerations, stresses, and finally increased wear of the engineering system, which might lead to catastrophic consequences. Moreover, analysis and prediction of the system's response and resistance become substantially more difficult when in most cases the underlying equations are unsolvable. On the other hand, the HCVI-oscillator can be used for engineering purposes such as vibration mitigation and to serve as a passive energy absorber. The latter is referred to as the hybrid cubic vibro-impact nonlinear energy sink (HCVI-NES).

The nonlinear PEAs, referred to as nonlinear energy sinks (NESs) are known for their vibration mitigation capabilities for wide frequency ranges under impulsive, periodic, and stochastic excitations [3, 4, 5]. They perform ideally when a transient resonance capture (TRC) [6] with the primary structure takes place. One example of an NES design is the cubic NES [6]. This design was proven as effective for vibration mitigation under excitation of various kinds and for suspension of undesired structural dynamical regimes, such as aeroelastic flutter [7]. The concept of NES was introduced to overcome the main shortcomings of the well-known linear PEA called tuned mass damper (TMD). The latter performs well only when sustained resonance takes place and the excitation is of relatively low magnitude, otherwise, it loses its linear properties. However, when the primary structure is exposed to low-energy excitations TRC is not achieved, and therefore the NES performs poorly. As a solution, the concept of the hybrid NES (HNES) was introduced [8]. The HNES includes the dynamical properties of multiple energy absorbers to allow improved vibration protection capabilities via a feature referred to as adaptive nonlinearity. In recent studies, it was shown that the dynamical richness of the HNES and coexistence of multiple dynamical regimes is essential for improved energy absorption capabilities for broader energy and frequency ranges [9, 10]. In other terms, the HNES takes advantage of the coexistence of multiple amplitude-dependant regimes to yield an adaptive nonlinearity that changes its dynamical properties in accordance with the magnitude and frequency of the external excitation. Often, the energy absorption capabilities of PEAs are evaluated by the maximal instantaneous energy measure captured by the absorber. Due to the transient nature of the TRC mechanism, this quantity cannot be evaluated based on the steady-state response of the system and thus, it heavily relies on numerical simulations. However, the existence of adaptive and non-smooth nonlinearity poses a mathematical challenge in describing and predicting the dynamical responses which are governed by transient phenomena under external disturbances. Hence, traditional asymptotic methods such as perturbation-based approaches (multiple scales, Lindstedt–Poincaré [11], complexification averaging [12, 13], harmonic balance [14]) become inapplicable. Moreover, those methods describe the response of forced systems in steady-state, mainly in conditions of primary resonance, and allow to obtain safety boundaries in the excitation space, but don't provide insights and understanding regarding the transient dynamics of the system. Additionally, the existence of vibro-impacts contributes to a hardening nonlinearity that leads to an increase in the response frequency. Therefore, most of the methods mentioned above are hardly applicable. In this paper, we focus on the analytical description of the transient dynamics of an HCVI-oscillator under an external harmonic forcing. More specifically, the hybrid oscillator is modeled by a classical particle in a one-dimensional hybridization of a quartic potential well and a square potential well with infinite depth, that correspond to the cubic nonlinearity and the collisions with rigid barriers, respectively. This hybrid equivalent potential well allows the coexistence of both SNOs due to cubic nonlinearity and the HCVI-regime due to the effect of the additional infinitely strong nonlinearity associated with collisions. The oscillator is initially at rest at the bottom of the well when the external force is switched on. It was shown in previous works [15, 16, 17, 18] that transient dynamical regimes can be accurately described using the reduced resonance manifold (RM) of the system that describes the slow dynamics of the system. Distinct dynamical regimes and their corresponding bifurcations are related to the topological modifications of a special phase trajectory on the RM that correspond to a set of initial conditions. The trajectory that corresponds to zero initial conditions is referred to as the limiting phase trajectory (LPT) [19, 20].

In the current work, we use canonical transformation to the action-angle (AA) variables and a proper averaging technique to describe all dynamical regimes that arise in the HCVI-oscillator and formulate the corresponding transition boundaries in the forcing parameters space. The main goal is to provide an analytical prediction of the particle's dynamical response and its maximal transient energy level (or absorption rate when thought of as a vibration mitigation solution or HCVI-NES) under monochromatic harmonic forcing. Moreover, we aim to describe all possible dynamical regimes that arise in the system, the boundaries associates with the transitions from one regime to another, and the underlying dynamical mechanisms that govern those transitions.

This paper is structured as follows: In Section 2 the dynamical model of the HCVI-oscillator is described. In Section 3 AA formalism is performed and the dynamical regimes are explained, the resonance manifold is computed and its structure is investigated. In Section 4 we describe the underlying bifurcation mechanisms that govern the transitions between dynamical regimes and energy levels and obtain the corresponding transition boundaries on the forcing parameters space. Frequency response curves are obtained, as well as a mapping between forcing parameters and the resulting maximal transient energy. Section 5 includes numerical validations of the analytic results. Section 6 is dedicated to the concluding remarks.

## 2. Model description

The system considered is the forced HCVI-oscillator with mass $m$ moving in a cavity of length $2d$, with elastic collisions with restitution coefficient $\kappa$ at both ends of the cavity, as shown in Fig. 1. The oscillator is subjected to a time-dependent forcing $\bar{F}(t) = F\sin(\bar{\Omega}t)$ and is attached to the rigid walls by a pair of cubic springs of stiffness $k/2$ each. The dimensional displacement of the oscillator with respect to its equilibrium position is denoted by $x$.

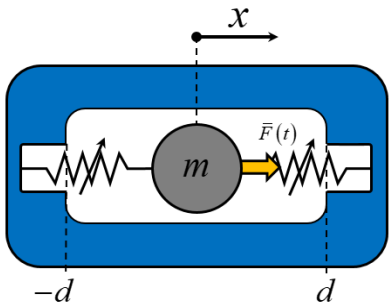

Figure 1: Scheme of the forced hybrid cubic vibro-impact (HCVI) oscillator

The normalized non-dimensional equation of motion of the system in shown in Eq. (1).

Here $q = x/d$ and $\tau = \bar{\omega}t$ are the non-dimensional displacement and time scale, respectively, where $\bar{\omega} = \sqrt{k/m}d/2$. The normalized forcing amplitude and frequency are $f = 4F/kd^3$ and $\Omega = \bar{\Omega}/\bar{\omega}$, respectively.

$$\ddot{q} + 4q^3 + \kappa \sum_j \dot{q}(\tau_j^-)\delta(\tau - \tau_j) = f\sin(\Omega\tau) \qquad (1)$$

Here $f$ and $\Omega$ are the non-dimensional forcing amplitude and frequency, respectively. Dot denote differentiation with respect to the non-dimensional time-scale $\tau$, and $\tau_j$ is the time instance of the $j^{th}$ collision since the forcing switched-on at $\tau = 0$. Parameter $\kappa$ is the restitution coefficient, and $\delta(\tau)$ is the Dirac delta function. We consider purely elastic collisions and take a restitution coefficient of unity $\kappa = 1$ accordingly. Thus, since the system is conservative, the motion of the HCVI-oscillator can be modeled by an equivalent potential well with infinite depth as shown in Eq. (2) and Fig. 2. This well is a hybridization of a quartic and a square potential well, associated with the cubic stiffness and the vibro-impact nonlinearity, respectively. Without loss of generality, in the further analysis, we consider $k = 4$. The transition between SNO and HCVI-regime corresponds to $|q| = 1$, i.e., instantaneous energy $E = 1$.

$$U(q) = \begin{cases} q^4 & , 0 < E \leq 1 \\ \infty & , E > 1 \end{cases} \qquad (2)$$

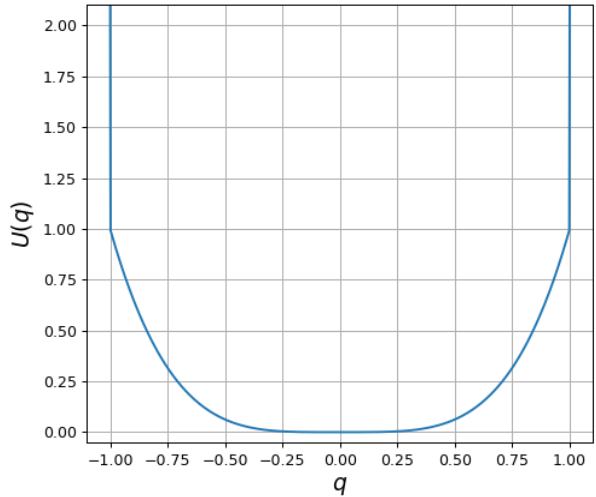

Figure 2: The equivalent potential well of the hybrid cubic vibro-impact (HCVI) oscillator. Energy values of $U(q) \in (0, 1)$ correspond to the regime of the SNO regime and values of $U(q) \in [1, \infty)$ correspond to the HCVI-regime. The limiting case of $U(q) = 1$ corresponds to a bifurcation of type I (dashed line), i.e. transition between one dynamical regime to another.

## 3. Computation of the resonance manifold

Following [21, 15], the dynamical system in Eq. (1) can be derived from the following Hamiltonian:

$$H = H_0(p, q) - fq\sin(\Omega\tau), \quad H_0(p, q) = \frac{1}{2}p^2 + q^4 \qquad (3)$$

Here $p = \dot{q}$ is the momentum of the HCVI-oscillator and the term $H_0(p, q)$ represents the free motion of the bounded oscillator, i.e. in absence of external forcing. Thus, the latter is also referred to as the conservative component of the Hamiltonian. The transformation to action-angle (AA) variables is conducted using the following well-known formulas [22]:

$$I(E) = \frac{1}{2\pi} \oint p(q, E)dq, \quad \theta = \frac{\partial}{\partial I} \int_0^q p(q, I)dq \qquad (4)$$

Here $I$ and $\theta$ are the action and angle variables, respectively. In Eq. (4), $H_0 = E$ defines a constant energy level. Explicit formulas for the canonical variables transformation $p(I, \theta)$ and $q(I, \theta)$, are obtained by inverting Eq. (4). The conservative component of the Hamiltonian depends only on the action variable: $H_0 = E(I)$. Then, the Hamiltonian of the system can be written in terms of AA variables as follows:

$$H = H_0(I) - fq(I, \theta)\sin(\Omega\tau) \qquad (5)$$

Due to $2\pi-$ periodicity of the angle variable $\theta$, Eq. (5) can be reframed using the following Fourier series [23]:

$$H = H_0(I) + \frac{if}{2} \sum_{n=-\infty}^{\infty} q_n(I) \left(e^{i(n\theta+\Omega\tau)} - e^{-i(n\theta-\Omega\tau)}\right), \quad q_n = \bar{q}_{-n} \qquad (6)$$

Here bar represents the complex conjugate. Then, the Hamilton equations take the following form:

$$\begin{aligned}
\dot{I} &= -\frac{\partial H}{\partial \theta} = \frac{f}{2} \sum_{n=-\infty}^{\infty} nq_n(I) \left(e^{i(n\theta+\Omega\tau)} - e^{i(n\theta-\Omega\tau)}\right) \\
\dot{\theta} &= \frac{\partial H}{\partial I} = \frac{\partial H_0}{\partial I} + \frac{if}{2} \sum_{n=-\infty}^{\infty} \frac{\partial q_n(I)}{\partial I} \left(e^{i(n\theta+\Omega\tau)} - e^{i(n\theta-\Omega\tau)}\right)
\end{aligned} \qquad (7)$$

In the current analysis, we consider vicinity of the frequency of the external forcing to one, i.e. $\Omega \approx 1$. Hence, we assume that the phase variable $\nu = \theta - \Omega\tau$ evolves slower than the higher harmonics $n > 1$. Then, we average Eq. (7) over the fast phase variables and obtain the following slow evolution equations:

$$\begin{aligned}
\dot{J} &= -\frac{f}{2}\left(q_1(J)e^{i\nu} + \bar{q}_1(J)e^{-i\nu}\right) \\
\dot{\nu} &= \frac{\partial H_0}{\partial J} - \frac{if}{2}\left(\frac{\partial q_1(J)}{\partial J}e^{i\nu} - \frac{\partial \bar{q}_1(J)}{\partial J}e^{-i\nu}\right) - \Omega
\end{aligned} \qquad (8)$$

Here $J(\tau) = \langle I(\tau) \rangle$ is the averaged action variable. Using Eq. (8), one can obtain the following conservation law:

$$C(J, \nu) = H_0(J) - \frac{if}{2}\left(q_1(J)e^{i\nu} - \bar{q}_1(J)e^{-i\nu}\right) - \Omega J \qquad (9)$$

Eq. (9) represents a set of resonance manifolds (RMs). Variable $C$ is determined by the initial conditions. In the current study we focus on the case in which the HCVI-oscillator begins from rest. The contour on the RM that correspond to this case, i.e. $H_0(J) = 0$, is called the limiting phase trajectory (LPT) [13, 12, 19, 20]. Now, the averaged action variable is given by the following expression:

$$J(\xi) = \begin{cases} \beta\xi^{\frac{3}{4}} & , \xi \in [0,1) \\ \frac{2\sqrt{2}}{\pi} {}_2\mathbf{F}_1\left(-\frac{1}{2}, \frac{1}{4}; \frac{5}{4}; \frac{1}{\xi}\right)\sqrt{\xi} & , \xi \in [1, \infty) \end{cases}, \quad \beta = \frac{4}{3\pi}\mathbf{K}\left(\frac{1}{\sqrt{2}}\right) \qquad (10)$$

Here, the averaged energy of the oscillator is denoted by $\xi(\tau) = \langle E(\tau) \rangle$ while averaging is performed over the fast phase variables. Function $\mathbf{K}$ is the elliptic integral of the first kind, and ${}_{\mathbf{p_1}}\mathbf{F}_{\mathbf{p_2}}$ is the generalized hypergeometric function. Detailed derivations of these expressions are presented in Appendix A. The relation between the averaged energy and frequency of the HCVI-oscillator is shown in Eq. (11) and Fig. 3.

$$\omega(\xi) = \left(\frac{\partial J}{\partial \xi}\right)^{-1} = \begin{cases} \frac{\pi}{\mathbf{K}\left(\frac{1}{\sqrt{2}}\right)}\xi^{\frac{1}{4}} & , \xi \in (0,1) \\ \frac{5\sqrt{2}\pi\xi^{\frac{3}{2}}}{10\xi\, {}_2\mathbf{F}_1\left(-\frac{1}{2},\frac{1}{4};\frac{5}{4};\frac{1}{\xi}\right)+2\, {}_2\mathbf{F}_1\left(\frac{1}{2},\frac{5}{4};\frac{9}{4};\frac{1}{\xi}\right)} & , \xi \in [1, \infty) \end{cases} \qquad (11)$$

Eq. (10)-(11) are valid for both the SNO-regime and HCVI-regime. As one can see, for $\xi \in (0,1)$ the frequency of the oscillator increases with energy. This is expected since the cubic term in Eq. (1) is positive, and thus contributing a hardening nonlinearity. Then, for $\xi > 1$ the gradient of the frequency increase dramatically due to the hardening effect of the impacts.

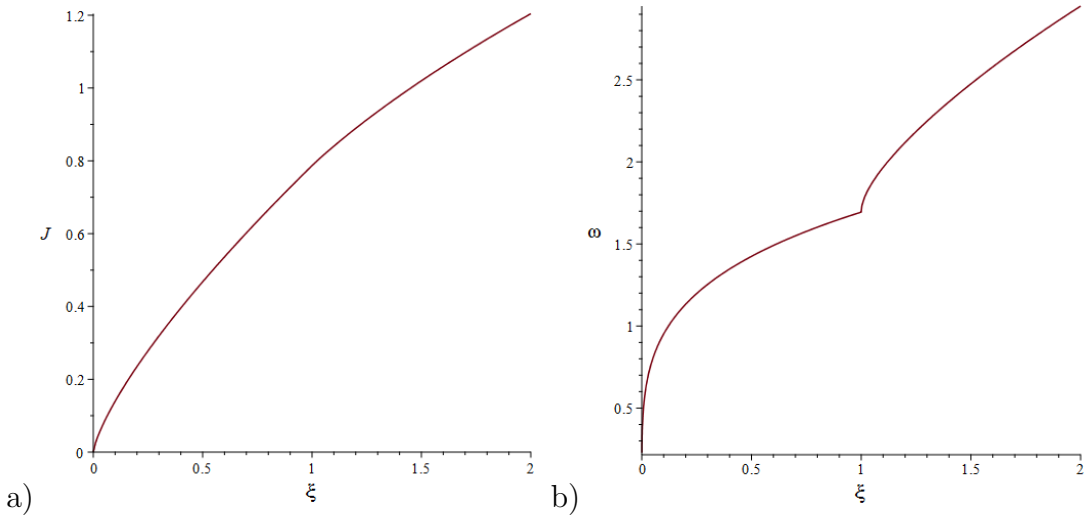

Figure 3: a) The averaged action, and b) the frequency of the HCVI-oscillator vs. the averaged energy $\xi$, according to Eq. (10) and Eq. (11), respectively. Ranges $\xi \in (0,1)$ and $\xi \geq 1$ correspond to the SNO-regime and the HCVI-regime, respectively. Monotonic increase of the frequency $\omega(\xi)$ stems from hardening non-linearities associated with both the positive cubic term and elastic impacts.

Following Eq. (4), the solution of the system is given in terms of AA variables in the following form:

$$q(\theta, \xi) = \xi^{\frac{1}{4}}\mathrm{sn}\left(\gamma(\xi)\theta, i\right) = a_0 + \sum_{n=1}^{\infty} a_n \sin(n\theta), \quad \gamma(\xi) = \frac{\sqrt{2}E^{\frac{1}{4}}}{\omega(E)} \qquad (12)$$

Here $\mathrm{sn}(x,i)$ is the Jacobi elliptic function of module $i = \sqrt{-1}$. Variable $\gamma(\xi)$ is plotted in Fig. 4. Full derivation is shown in Appendix A. By following Eq. (9) and Eq. (12), and recalling that $q_1 = -ia_1/2$, we calculate the coefficient of the first term of the Fourier series of the displacement, i.e. $a_1$. Coefficient $a_1$ is written in Eq. (13) and plotted in Fig. 5.

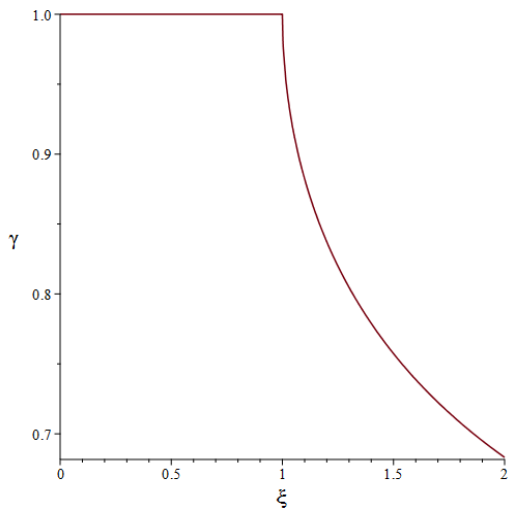

Figure 4: Function $\gamma(\xi)$ vs. the averaged energy of the HCVI-oscillator.

$$a_1(\xi) = \begin{cases} \alpha \xi^{\frac{1}{4}} & \xi \in [0,1) \\ \frac{2\alpha \sin(\pi\gamma)}{\pi(1-\gamma^2)} \xi^{\frac{1}{4}} & \xi \in [1,\infty) \end{cases}, \quad \alpha = \frac{2\pi\eta}{\mathbf{K}(i)(1+\eta^2)} \qquad (13)$$

Here, $\eta = e^{-\pi/2}$. The non-smoothnesses at $\xi = 1$ in Fig. 4 and Fig. 5 stem from the non-differentiability of the response frequency of the oscillator $\omega(\xi)$. Since Eq. (10) is not invertible, the averaged energy cannot be explicitly expressed by the averaged action. Therefore, we parameterize all values following from the AA transformation by the averaged energy $\xi(\tau)$ instead of the averaged action $J(\tau)$. As a result, the conservation law takes the following form:

$$C(\nu, \xi) = \xi - \frac{f}{2} a_1(\xi) \cos(\nu) - \Omega J \qquad (14)$$

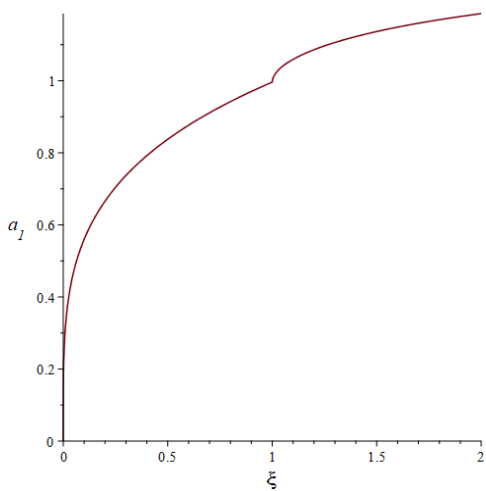

Figure 5: The first Fourier coefficient of $q(\theta, \xi)$ versus the averaged energy of the HCVI-oscillator, $a_1(\xi)$.

As mentioned, in the current work we focus on the case of zero initial conditions. Thus, the response of the system is represented by the level line on the phase cylinder $\nu \in [0, 2\pi]$, $\xi \in [0, \infty]$, that corresponds to $C(\nu, \xi) = 0$, i.e. the LPT. The maximal transient averaged energy level reached by the HCVI-oscillator $\tilde{\xi}(\Omega, f)$ corresponds to the maximal height on the phase cylinder to which the LPT reaches from the bottom of the cylinder $\xi = 0$.

Now, let us explore the topological structure (phase portrait) of the RM, defined by the conservation law in Eq. (14). We assume that the oscillator is captured to the RM throughout its motion. The stationary points of the RM correspond to the following equations:

$$\frac{\partial C(\nu, \xi)}{\partial \nu} = 0, \quad \frac{\partial C(\nu, \xi)}{\partial \xi} = 0 \qquad (15)$$

From the first equation in Eq. (15) we conclude that the stationary points can be only on lines $\nu_0 = 0, \pi$ on the phase cylinder. The latter corresponds to the saddle point of the RM. In the following section, we explain and explore the relationship between the RM structure and the underlying dynamical mechanisms that correspond to gradual/drastic increase/decrease of the response energy of the HCVI-oscillator (or absorption rate for energy absorption applications).

## 4. Bifurcations and underlying mechanisms

In this section, the dynamical mechanisms that govern the bifurcations and transitions between different dynamical regimes are identified and analyzed both analytically and numerically. Let us define two distinct bifurcation types. The former is associated with the transition between SNO to HCVI regime and vice versa, and the latter with the transient crossing of a chosen energy level. Thus, the bifurcation of type I is a particular case of bifurcation of type II for which the energy threshold equals one, i.e. $\tilde{\xi} = 1$. Both bifurcation types correspond to a set of critical forcing amplitudes and frequencies. Those sets of values create contours in the plane of forcing parameters, which are referred to as the transition boundaries. The main goal of the current analysis is to describe those transition boundaries analytically for any given maximal transient energy level $\tilde{\xi}$. In physical and engineering systems, this value can serve as a measure of structural damage, while for vibration protection applications it serves as a measure of the energy absorption rate and thus the mitigation capabilities of the PEA [24, 8].

In previous works [25, 26] it was demonstrated that the transition boundary curves (also referred to as escape curves when associated with the escape of a classical particle from a potential well) share a common geometric property of a sharp minimum. Recently it was shown that this 'dip' corresponds to the intersection of two contours that corresponds to two competing dynamical mechanisms. The former is associates with a gradual increase of the amplitude, where bifurcation occurs when it reaches the critical value. In this case, the LPT reaches the bifurcation value or maximal threshold and is thus referred to as the 'maximum mechanism' (MM). The second mechanism is associated with an abrupt jump of the response amplitude and corresponds to the passage of the LPT through the saddle point of the RM followed by a sudden increase of the response energy towards the critical energy threshold. Therefore, the latter is referred to as the 'saddle mechanism' (SM). Due to the abrupt nature of the 'jump' phenomenon, the SM is considered more potentially hazardous for engineering applications, and on the other hand, a very effective energy absorption mechanism that can be utilized by PEAs, allowing extensive broadband vibration protection for sensitive engineering systems.

## 4.1. Regular and degenerate saddle mechanism

Following Eq. (14)-(15), the bifurcation through the SM corresponds to the following relations:

$$\begin{cases} C(\nu = \pi, \xi_s | f_s) = \xi_s + \frac{f_s}{2} a_1(\xi_s) - \Omega J(\xi_s) = 0 \\ \frac{\partial C}{\partial \xi}(\nu = \pi, \xi_s | f_s) = 1 + \frac{f_s}{2} a_1'(\xi_s) - \Omega J'(\xi_s) = 0 \end{cases} \quad (16)$$

Here $\xi_s$ is the energy level of the saddle point of the RM, and $f_s$ is the critical forcing amplitude associated with a bifurcation through the SM. Tag stands for differentiation with respect to the averaged energy $\xi$, i.e. $(\ )' \equiv \partial/\partial\xi$. When bifurcation of type II through the SM takes place, the LPT reaches the saddle point of the RM at $\xi_s$ and than jumps to the energy boundary that corresponds to $\tilde{\xi}$. An implicit relation between the forcing frequency and the energy level of the saddle point is obtained by merging both equations in Eq. (16) while eliminating the critical forcing amplitude:

$$\Omega = \frac{a_1(\xi_s) - a_1'(\xi_s)\xi_s}{a_1(\xi_s)J'(\xi_s) - a_1'(\xi_s)J(\xi_s)} \quad (17)$$

The relation between the forcing frequency $\Omega$ and the resulting saddle energy level $\xi_s$ Eq. (17) is plotted numerically in Fig. 6. This plot shows the values of all stationary points located at line $\nu = \pi$ on the RM vs. the forcing frequency $\Omega$. One can see in Fig. 6 that for increasing forcing frequency values, the energy level of the saddle point shifts from zero towards one, where it turns into a degenerate saddle point at $\hat{\Omega} = 1.9062$ (corresponds to substituting $\xi_s = 1$ in Eq. (17)). For $\Omega = 2.5978$ a regular saddle and a minimum point emerge. This birth of a pair of stationary points corresponds to Eq. (18) which is obtained by nullifying the derivative of Eq. (17) with respect to the $\xi$. For $\Omega > 2.5978$, the SM is governed by the regular saddle point rather than the degenerate saddle point. Generally, even though the degenerate saddle point exists for any value of forcing frequency it becomes significant only in absence of the regular saddle point. The vertical asymptote associated with $\xi = 1.01989$ corresponds to the singularity of Eq. (17), i.e. to the energy level that eliminates its denominator.

$$a_1''(J - \xi J') - J''(a_1 - \xi a_1') = 0 \quad (18)$$

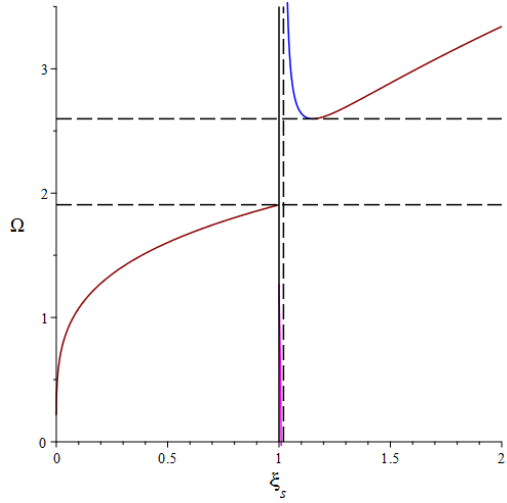

Figure 6: Plot of the stationary points of the RM in the energy-frequency plane according to the implicit expression Eq. (17); saddle points, degenerate saddles, degenerate stationary points, and maxima are colored in red, black, magenta, and blue, respectively. The black dashed vertical line correspond to the asymptote $\xi = 1.01989$ (Eq. (18)). The upper and lower horizontal dashed black lines corresponds to $\Omega = \hat{\Omega} = 1.9062$ (corresponds to substituting $\xi_s = 1$ in Eq. (17)) and $\Omega = 2.5978$ (obtained by eliminating the derivative of Eq. (17) with respect to the $\xi$), respectively.

Since the assumptions of the current analytical treatment are valid only in vicinity of $\Omega = 1$, further analysis overlooks the emergence of new saddle point for $\Omega > 2.5978$. Hence, the energy level associated with the regular and degenerate saddle points is taken as follows:

$$\xi_s = \begin{cases} \left(\frac{2\beta}{3}\right)^4 \Omega^4 & ,\Omega \in (0,\hat{\Omega}) \\ 1 & ,\Omega \geq \hat{\Omega} \end{cases}, \quad \xi_s = 1 \to \hat{\Omega} = \frac{3}{2\beta} \approx 1.9062, \quad \beta = \frac{4}{3\pi}\mathbf{K}\left(\frac{1}{\sqrt{2}}\right) \quad (19)$$

Substituting Eq. (17) into the first equation in Eq. (16), obtains the following expression for the critical forcing amplitude associated with bifurcation through the regular and degenerate SM:

$$f_s(\Omega) = \frac{2}{a_1(\xi_s)}\left(\Omega J(\xi_s) - \xi_s\right) = \begin{cases} \frac{8\beta^3}{27\alpha}\Omega^3 & ,\Omega \in (0,\hat{\Omega}) \\ \frac{2}{\alpha}(\Omega\beta - 1) & ,\Omega > \hat{\Omega} \end{cases} \quad (20)$$

Transition from the regular SM to the DSM corresponds to the following excitation amplitude:

$$\hat{f} = f_s(\hat{\Omega}) = \frac{1}{\alpha} \approx 1.0471 \quad (21)$$

As one can learn from Eq. (20), in contrast to the MM, the critical forcing amplitude associates with bifurcation through the SM is independent of the threshold transient energy level $\tilde{\xi}$. Demonstration of the bifurcation of type II for $\tilde{\xi} = 1.5$ (chosen arbitrarily) through the regular and degenerate SM from the perspective of the phase portrait is shown in Fig. 7-Fig. 8.

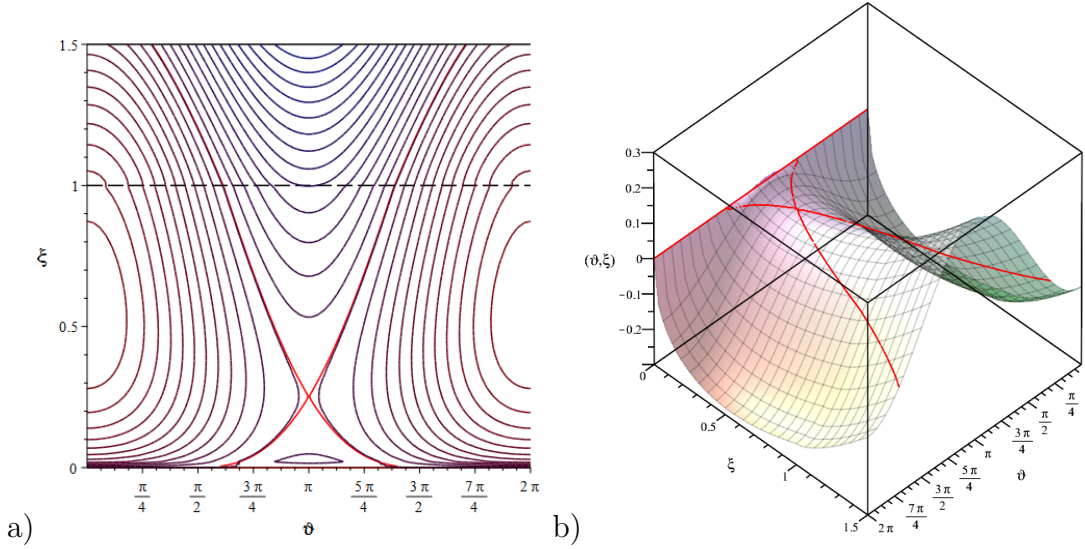

Figure 7: Bifurcation of type II for $\tilde{\xi} = 1.5$ through the regular saddle mechanism in phase portrait, defined by $C(\nu, \xi)$. The LPT is marked by a red line. Dashed and solid black horizontal lines correspond to $\xi = 1$ and $\tilde{\xi} = 1.5$, respectively. The former and the latter correspond to bifurcations of type I and II, respectively. For $\Omega = 1.35, f = 0.35655$, and a) 2D projection of $C(\nu, \xi)$ on the $(\nu, \xi)$ plane, b) 3D plot of $C(\nu, \xi)$.

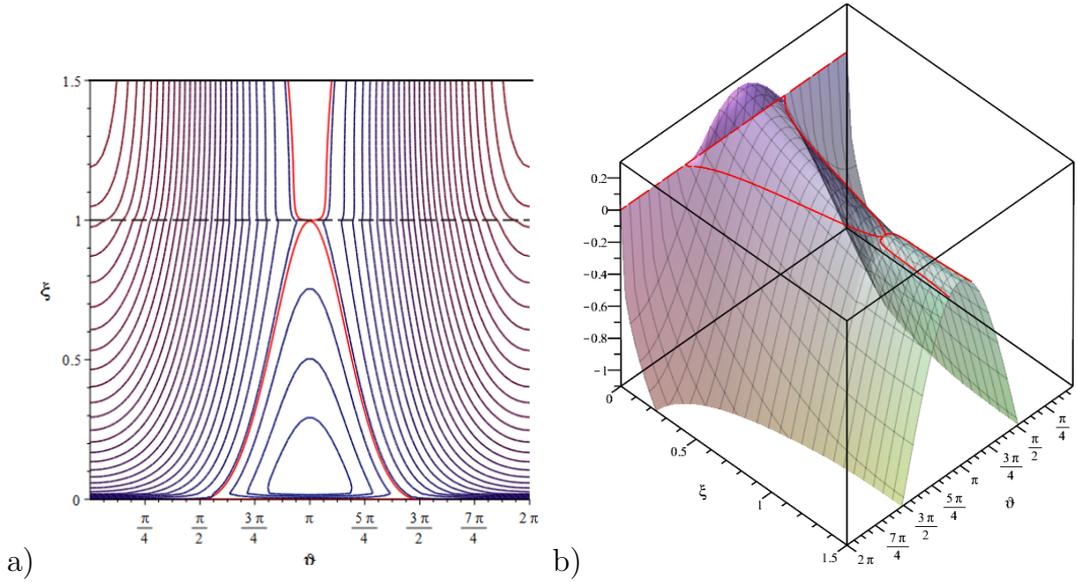

Figure 8: Bifurcation of type II for $\tilde{\xi} = 1.5$ through the degenerate saddle mechanism in phase portrait, defined by $C(\nu, \xi)$. The LPT is marked by a red line. Dashed and solid black horizontal lines correspond to $\xi = 1$ and $\tilde{\xi} = 1.5$, respectively. The former and the latter correspond to bifurcations of type I and II, respectively. For $\Omega = \hat{\Omega} = 1.9062, f = 1.0471$, and a) 2D projection of $C(\nu, \xi)$ on the $(\nu, \xi)$ plane, b) 3D plot of $C(\nu, \xi)$.

### 4.2. Maximum mechanism at $\nu = 0$

For $f > f_s(\Omega)$, bifurcations are dominated by the maximum mechanism (MM) at $\nu = 0$. In this scenario, the LPT directly approaches the critical energy level $\tilde{\xi}$ at $\nu = 0$ (bifurcation of type II for general $\tilde{\xi}$, bifurcation of type I for $\tilde{\xi} = 1$), without passing thought the saddle point at $\nu = \pi$. Thus, this MM is referred to as the MM0. The corresponding critical forcing amplitude is obtained from Eq. (15) using the following set of equations:

$$\begin{cases} C(\nu=0,\tilde{\xi}|f_{m,0}) = \tilde{\xi} - \frac{f_{m,0}}{2}a_1(\tilde{\xi}) - \Omega J(\tilde{\xi}) = 0 \\ \frac{\partial C}{\partial \xi}(\nu=0,\tilde{\xi}|f_{m,0}) = 1 - \frac{f_{m,0}}{2}\frac{\partial a_1(\tilde{\xi})}{\partial \xi} - \Omega \frac{\partial J(\tilde{\xi})}{\partial \xi} = 0 \end{cases} \quad (22)$$

The relations in Eq. (22) correspond to the fact that the LPT is tangent to the upper bound of the phase cylinder $\tilde{\xi}$ at $\nu = 0$. Solving Eq. (22) yields the following relation between the excitation frequency to the critical forcing amplitude:

$$f_{m,0}(\Omega|\tilde{\xi}) = \frac{2}{a_1(\tilde{\xi})}\left(\tilde{\xi} - \Omega J(\tilde{\xi})\right) \quad (23)$$

Bifurcation of type II for for $\tilde{\xi} = 1.5$ thought the MM0 is demonstrated graphically in Fig. 9 from the perspective of the phase portrait. As mentioned above, the MM0 is dominant only above the curve described in Eq. (20), i.e. for $f > f_s(\Omega)$.

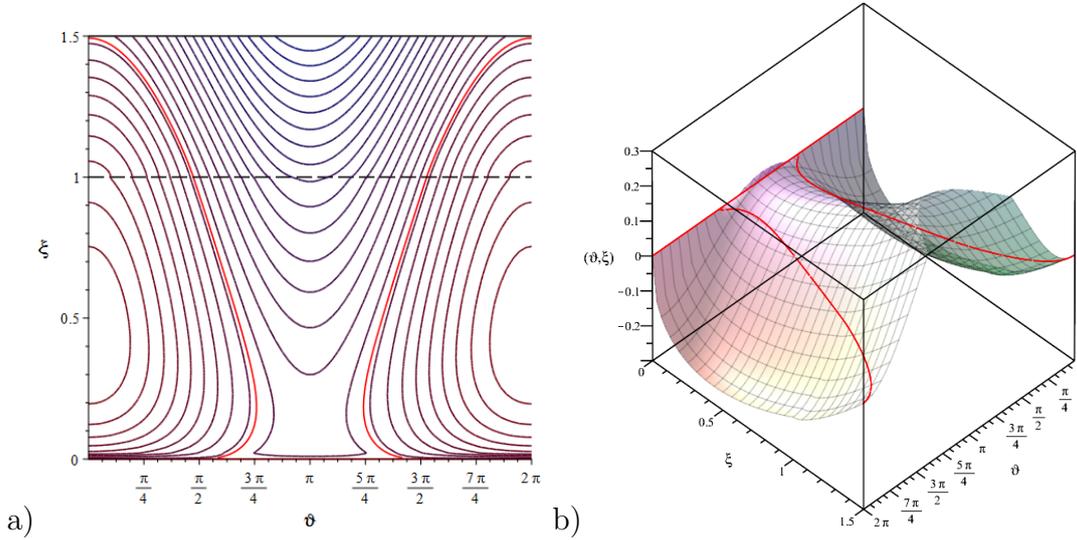

Figure 9: Bifurcation of type II for $\tilde{\xi} = 1.5$ through the maximum mechanism at $\nu = 0$ in the phase portrait, defined by $C(\nu, \xi)$. The LPT is marked by a red line. Dashed and solid black horizontal lines correspond to $\xi = 1$ and $\tilde{\xi} = 1.5$, respectively. The former and the latter correspond to bifurcations of type I and II, respectively. For $\Omega = 1.25$, $f = 0.39$, and a) 2D projection of $C(\nu, \xi)$ on the $(\nu, \xi)$ plane, b) 3D plot of $C(\nu, \xi)$.

4.3. Maximum mechanism at $\nu = \pi$

In contrast to the SM, for $f < f_s(\Omega)$, the upper and the lower branches of the LPT are yet to intersect since the forcing amplitude is insufficient for saddle bifurcation to take place. In this case, a sub-saddle MM takes place below the saddle point at $\nu = \pi$, i.e. for $\tilde{\xi} < \xi_s(\Omega)$. Thus, the latter is referred to as the MM$\pi$. In this case the critical excitation amplitude is smaller that the one associated with SM, and therefore the corresponding energy level in yet reached $\tilde{\xi} < \xi_s \leq 1$ and only SNO-regime can take place. This mechanism is referred to as the MM$\pi$, and it satisfies the following set of equations:

$$\begin{cases} C(\nu=\pi,\tilde{\xi}|f_{m,\pi}) = \tilde{\xi} + \frac{f_{m,\pi}}{2}a_1(\tilde{\xi}) - \Omega J(\tilde{\xi}) = 0 \\ \frac{\partial C}{\partial \xi}(\nu=\pi,\tilde{\xi}|f_{m,\pi}) = 1 + \frac{f_{m,\pi}}{2}\frac{\partial a_1(\tilde{\xi})}{\partial \xi} - \Omega \frac{\partial J(\tilde{\xi})}{\partial \xi} = 0 \end{cases} \quad (24)$$

The relations in Eq. (24) correspond to the fact that the LPT is tangent to the line $\tilde{\xi} < \xi_s(\Omega)$ on the phase cylinder at $\nu = \pi$. Solving Eq. (22) yields the following relation between the excitation frequency to the critical forcing amplitude:

$$f_{m,\pi}(\Omega|\tilde{\xi}) = -\frac{2}{a_1(\tilde{\xi})}\left(\tilde{\xi} - \Omega J(\tilde{\xi})\right) = -\frac{2}{\alpha}\left(\tilde{\xi}^{\frac{3}{4}} - \beta\Omega\tilde{\xi}^{\frac{1}{2}}\right) \quad (25)$$

Type I bifurcation (escape) thought the MM$\pi$ is demonstrated graphically in Fig. 10.

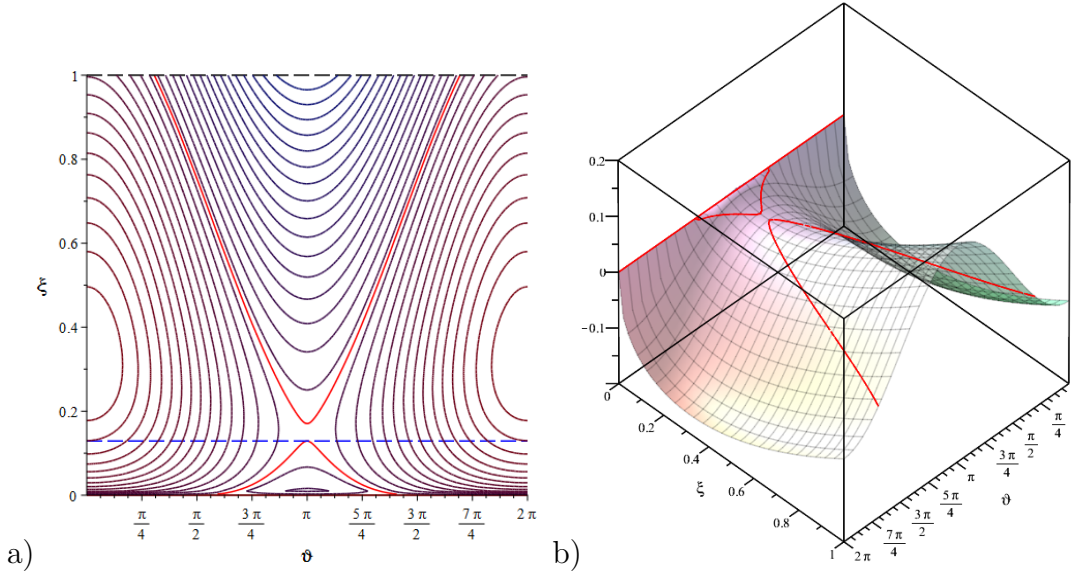

Figure 10: Escape through the maximum mechanism at $\nu = \pi$ (MM$\pi$) in the phase portrait, defined by $C(\nu,\xi)$. The LPT is marked by a red line. For $\Omega = 1.184, f = 0.25$. Dashed black and blue lines correspond to the maximal transient energy level associated with bifurcation of type I. i.e. $\tilde{\xi} = 1$, and to the maximal energy level reached by the LPT, i.e. $\tilde{\xi} = 0.13$ (satisfies $\tilde{\xi} < \xi_s(\Omega = 1.184) = 0.1488$); a) 2D projection of $C(\nu,\xi)$ on the $(\nu,\xi)$ plane, b) 3D plot of $C(\nu,\xi)$.

Coexistence of both the MM0 and the SM is achieved by satisfying both Eq. (23) and Eq. (20) simultaneously, i.e. for $f_m(\Omega^*|\tilde{\xi}) = f_s(\Omega^*)$, where $\Omega^*$ corresponds to the intersection point of both branches on the forcing parameters plane $\Omega - f$ (Eq. (26)). The RM structure that corresponds to this case is shown shown in Fig. 11.

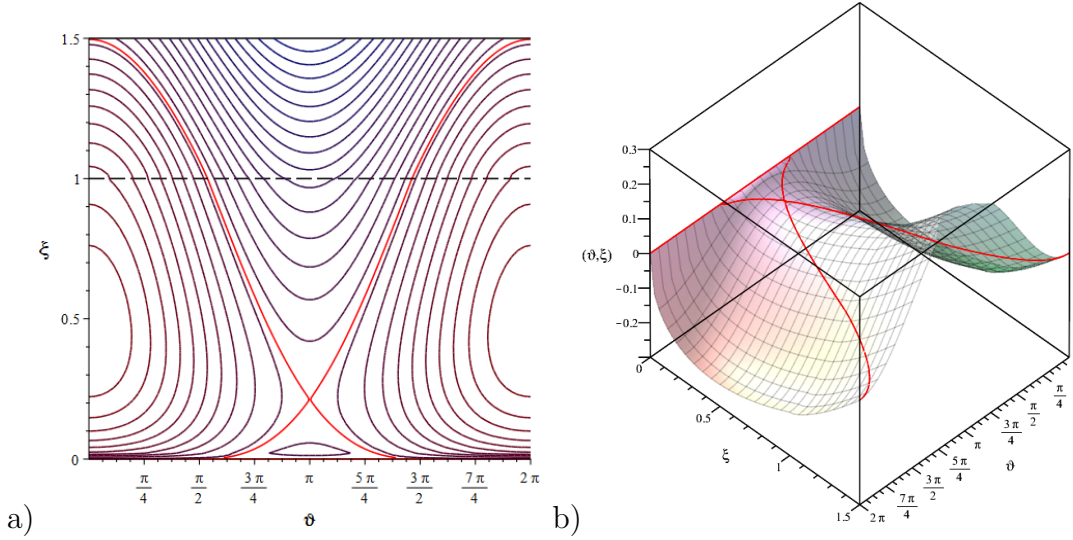

Figure 11: Bifurcation of type II for $\tilde{\xi} = 1.5$ through the coexistence of both the saddle mechanism (SM) and the maximum mechanism at $\nu = 0$ (MM0) in phase portrait, defined by $C(\nu, \xi)$. The LPT is marked by a red line. Dashed and solid black horizontal lines correspond to $\tilde{\xi} = 1$ and $\tilde{\xi} = 1.5$, respectively. The former and the latter correspond to bifurcations of type I and II, respectively. For $\Omega = 1.294$, $f = 0.314$, and a) 2D projection of $C(\nu, \xi)$ on the $(\nu, \xi)$ plane, b) 3D plot of $C(\nu, \xi)$.

$$\Omega^* = \frac{a_1(\xi_s^*)\tilde{\xi} + a_1(\tilde{\xi})\xi_s}{a_1(\tilde{\xi})J(\xi_s^*) + a_1(\xi_s^*)J(\tilde{\xi})} \qquad (26)$$

Here $\xi_s^* \equiv \xi_s(\Omega^*)$ in accordance to Eq. (17). The transition boundaries that correspond to bifurcation of type I ($\tilde{\xi} = 1$) are shown in Fig. 12. As one can see, the blue and black branches correspond to bifurcations through the MM0 and MM$\pi$, respectively. The red branch corresponds to bifurcation through the SM. As one can see, the bifurcation of type I through the MM$\pi$ is overlaps the branch associated with the SM for $\Omega > \hat{\Omega}$. All three mechanisms correspond to three curves on the forcing parameters plane. The curve defined by $f(\Omega) = \max\{f_{m,0}(\Omega), f_s(\Omega), f_{m,\pi}(\Omega)\}$ and shown in Fig.12, is a contour with identical maximal transient energy, or an iso-energy level line. In other terms, for any excitation frequency, the upper branch among the triplet corresponds to the overruling mechanism that governs the transient process. Surprisingly, as one can see in Fig. 12, the universal property of a sharp minimum of the escape curve exists also in the current case, when the potential well is purely and strongly nonlinear and lacks a linear term. The right shift of the dip corresponds to the hardening effect of the quartic nonlinearity, i.e. the positive cubic term in the equation of motion (Eq. (1)). For larger values of critical transient energy level $\tilde{\xi} > 1$, the corresponding iso-energy curves would shift to the right while its minimum stays on the $f_s(\Omega)$ branch, and vise versa for lower $\tilde{\xi}$ values, as one can see in 13.

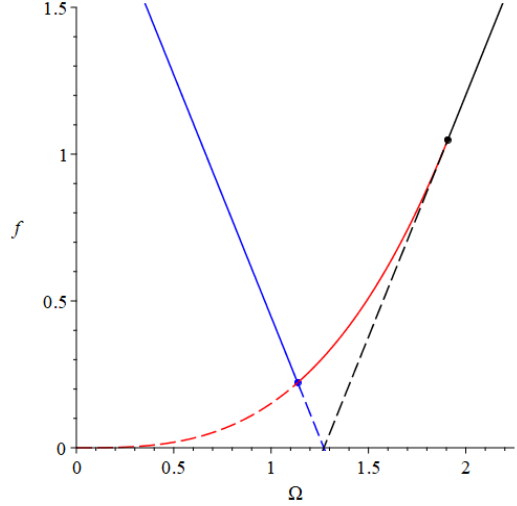

Figure 12: Transition boundary of a HCVI-oscillator. The blue and black branches correspond to escape through the MM0 and the MM$\pi$, respectively. The red line corresponds to escape through the saddle mechanism. The coexistence of the MM0 and SM is denoted by a blue circle, and the coexistence of the MM$\pi$ and the SM is denoted by a black circle. A dashed line means that the dynamical mechanism is overruled by another mechanism.

We use Eq. (23)-(20) to implicitly plot $\tilde{\xi}(\Omega, f)$- the inverse function of Eq. (23) that maps the forcing parameters to the resulting maximal transient response energy obtained by the HCVI-oscillator. Level lines $\tilde{\xi}(\Omega, f) = $ const represent iso-energy contours over the forcing parameters plane $\Omega - f$ for arbitrary transient energy level $\tilde{\xi}$. The energy jumps associated with bifurcation through the SM are demonstrated by the cliff shown in Fig. 13. This representation gives a full perspective on the HCVI-oscillator's response regimes, response energy levels, and energy absorption capabilities of the HCVI-NES over the entire parameters space of monochromatic harmonic excitations.

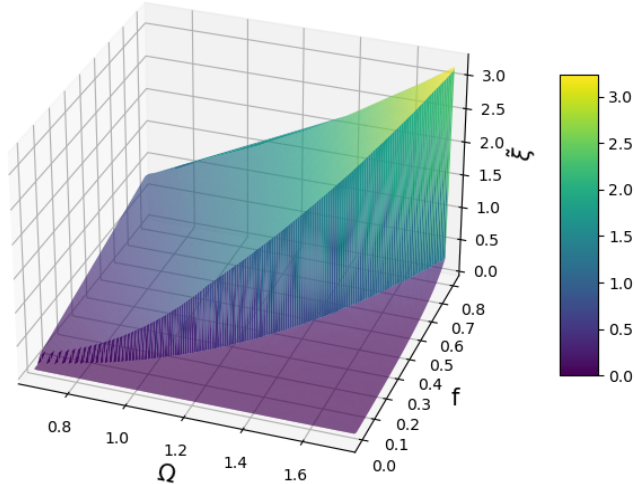

Figure 13: Plot of the maximal response energy $\tilde{\xi}(\Omega, f)$ over the forcing parameters plane.

#### 4.3.1. Frequency response

Since the amplitude of the oscillations is bounded by the rigid barriers at $|q| = 1$, the frequency response curve cannot be expressed in terms of the response amplitude for transient energy levels greater than one, i.e. beyond the type I bifurcation ($\tilde{\xi} > 1$). Therefore, the frequency response curves are defined in terms of maximal transient energy level of the oscillator's response rather than its steady-state response amplitude. The latter is defined as the intersection curve between the manifold $\tilde{\xi}(\Omega, f)$ and a plane that corresponds to a desired forcing amplitude $f = f_m$. Hence, the energy-based response curve of the HCVI-oscillator corresponds to the following expression, and plotted in Fig. 14:

$$\tilde{\xi}(\Omega, f = f_m) \Leftrightarrow f - f_m(\Omega|\tilde{\xi}) = 0 \qquad (27)$$

Here, $f_m(\Omega|\tilde{\xi})$ is taken from Eq. (23). The left and right branches of each frequency response curve in Fig. 14 correspond to type II bifurcation through the MM0 and the MM$\pi$, respectively. The horizontal dashed line corresponds to type I bifurcation (transition from SNO to HVI-regime and vise versa, $\tilde{\xi} = 1$), and the vertical dashed line corresponds to the sudden energy jump associated with a type II bifurcation through the SM, and thus take place for forcing frequency that corresponds to the inverse relation of $f = f_s(\Omega)$.

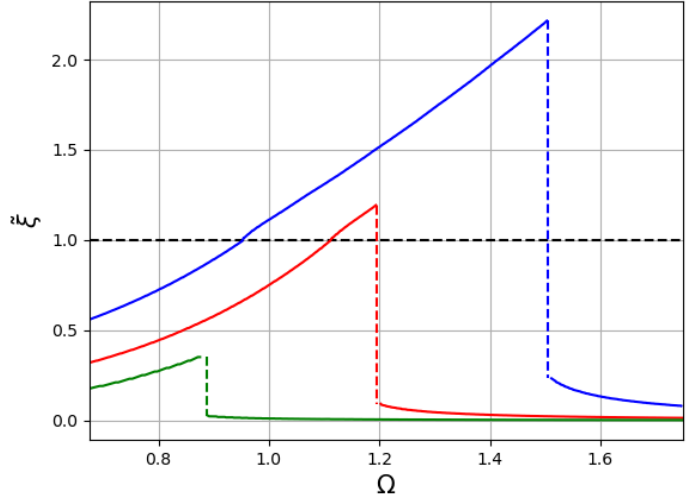

Figure 14: Frequency response curves for $f = 0.1, 0.25, 0.5$, colored in green, red, and blue, respectively. Vertical dashed lines correspond to non-smooth energy jumps associated with crossing the transition boundary through the saddle mechanism. Horizontal black dashed line corresponds to the transition between the SNO to HCVI regimes, i.e. bifurcation of type I, $\tilde{\xi} = 1$.

## 5. Numerical verification

Further exploration verifies the analytical results using numerical integration of the system's equation of motion, Eq. (1). First, we verify the analytical predictions for the occurrence of bifurcation of type I through the MM0. In Fig. 15 verification is performed in the time domain from the perspective of both displacement and energy. As one can see, type I bifurcation takes place when both and the displacement and the energy of the HCVI-oscillator reach one simultaneously. The latter corresponds to the transition from SNO to the HCVI-regime. In Fig. 16, type II bifurcation ($\tilde{\xi} = 1.5$) through the MM0 is demonstrated. There, for a gradual increase in the forcing amplitude $f$, there is a resulting gradual increase in the response energy, until the maximal transient energy reteaches the critical value of $\tilde{\xi} = 1.5$. On the other hand, in Fig. 17 a type II bifurcation ($\tilde{\xi} = 1.5$) through the SM is demonstrated. There, one can see, a sudden jump takes place from the saddle point at $\xi_s = 0.251$ (Eq. (17), Fig. 6) to the critical energy level $\tilde{\xi} = 1.5$ around time $\tau \approx 80$. At the same time, also a type I bifurcation occurs when the transition from SNO to HCVI-regime takes place by crossing energy level $\tilde{\xi} = 1$. The coexistence of both mechanisms is shown in Fig. 18, where the sadden jump through the saddle point lands exactly at the critical energy level, $\tilde{\xi} = 1.5$, in contrast to Fig. 17. Multiple sets of forcing frequencies and their critical forcing amplitudes are compared with theoretical results in Fig. 19(a). As one can see, there is a good agreement between the analytical and numerical results. This agreement deteriorates as the forcing frequency differs more significantly from one, where the assumptions considered in the analytical treatment lose validity.

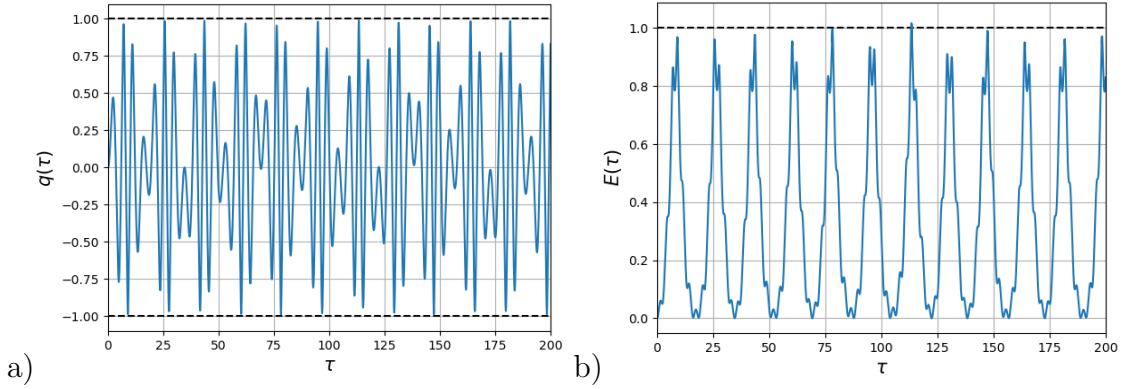

Figure 15: Bifurcation of type I ($\tilde{\xi} = 1$) through the maximum mechanism, for $\Omega = 1.0$ i.e. $f = 0.35$, a) displacement response, b) energy response. Dashed black lines correspond to the bifurcation energy value, i.e. $\tilde{\xi} = 1.0$.

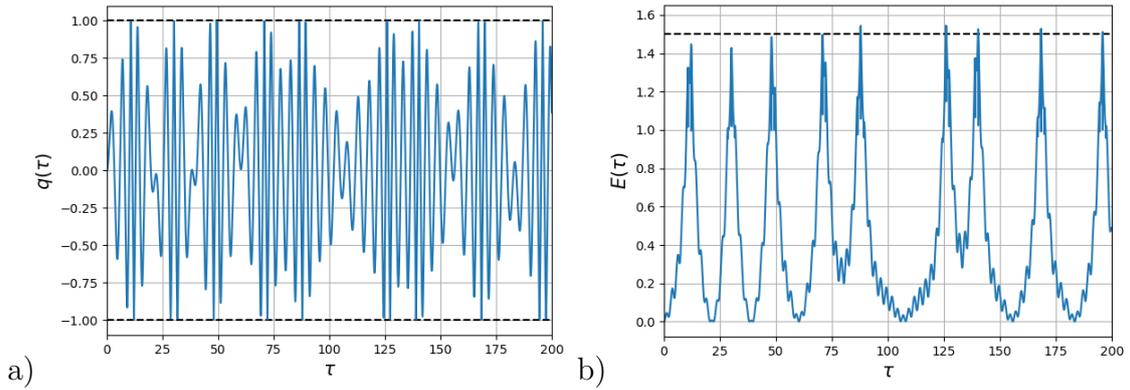

Figure 16: Bifurcation of type II ($\tilde{\xi} = 1.5$) through the maximum mechanism, for $\Omega = 1.25$ i.e. $f = 0.38$, a) displacement response, b) energy response. Dashed black lines correspond to the bifurcation energy value, i.e. $\tilde{\xi} = 1.5$.

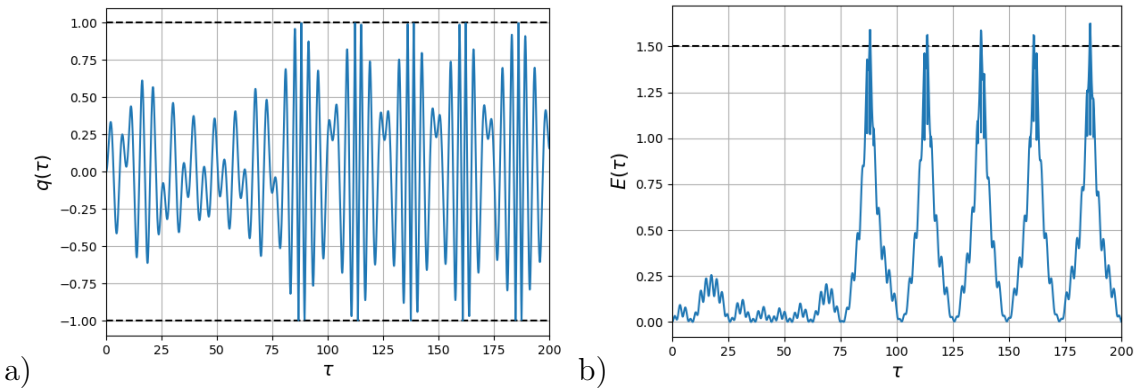

Figure 17: Bifurcation of type II ($\tilde{\xi} = 1.5$) through the saddle mechanism, for $\Omega = 1.35$ i.e. $f = 0.348$, a) displacement response, b) energy response. Dashed black lines correspond to the bifurcation energy value, i.e. $\tilde{\xi} = 1.5$.

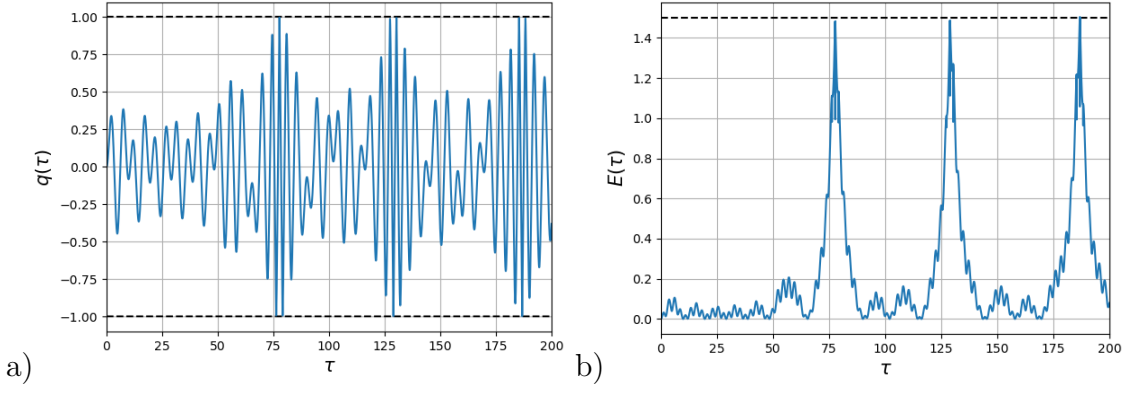

Figure 18: Bifurcation of type II ($\tilde{\xi} = 1.5$) through the coexistence of both the maximum and the saddle mechanisms, for $\Omega = 1.294$ i.e. $f = 0.32$, a) displacement response, b) energy response. Dashed black lines correspond to the bifurcation energy value, i.e. $\tilde{\xi} = 1.5$.

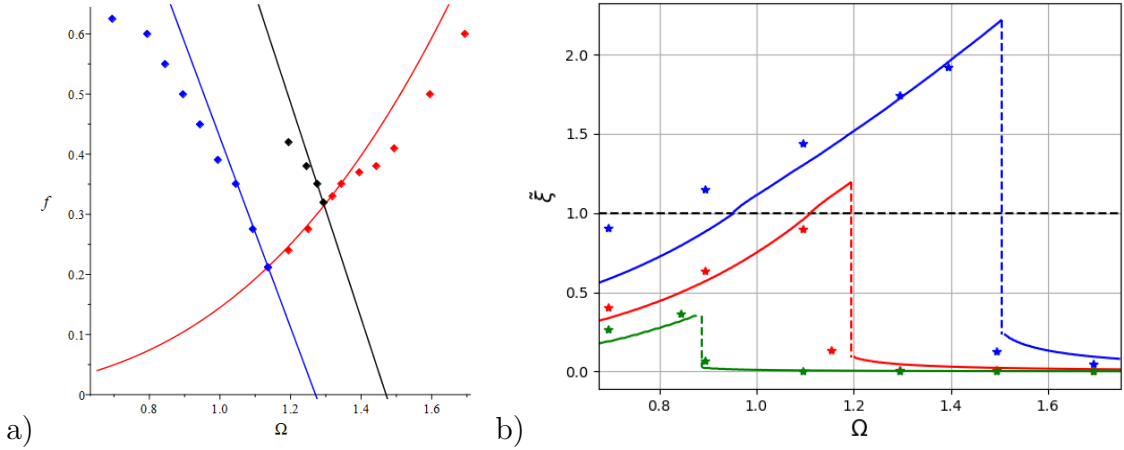

Figure 19: Numerical validation of the analytical results by integration of the equation on motion Eq. (1); a) bifurcation boundaries via MM0 (blue curve- bifurcation type I, black curve- bifurcation type II for $\tilde{\xi} = 1.5$) and SM (red curve). Corresponding numerical validations are marked with dots; b) energy-based frequency response curves for $f = 0.1, 0.25, 0.5$, colored in green, red, and blue, respectively.

## 6. Concluding remarks

The dynamical regimes that emerge in the hybrid cubic vibro-impact oscillator and NES and their underlying leading mechanisms were identified and explored analytically, to yield a full understating regarding the desired or potentially hazardous operational conditions. The main result of the current work was the formulating the analytical relation between the properties of a periodic monochromatic excitation and the resulting maximal transient energy level. Since the HCVI-oscillator is a strongly non-linear system that involves collisions and vibro-impact motion, traditional perturbation-based methods are inapplicable for the aforementioned task. Hence, canonical formalism was used to describe the dynamics in terms of action-angle variables. The slow-flow dynamics was described by a reduced cylindrical resonance manifold, and two types of bifurcations were identified. The former is associated with the transition between the dynamical regimes of smooth nonlinear oscillations and the hybrid cubic vibro-impact regime, and vice versa. The latter is associated with a chosen critical maximal transient energy level. Both bifurcations types correspond to transition boundaries in the plane of excitation parameters, and they overlap when the critical energy value equals the energy level that corresponds to the birth of the HVI-regime, i.e., $\tilde{\xi} = 1$. It was shown that the sharp minimum in the transition boundary curves is associated with the coexistence of two topological mechanisms of the limiting phase trajectory: the maximum mechanism at line $\nu = 0$ on the resonance manifold and the saddle mechanism. This pattern is universal for escape problems under periodic forcing. In contrast to previous studies, another dynamical mechanism that governs the transient increase in the system's energy level was identified: the maximum mechanism at $\nu = \pi$. Frequency response curves in terms of the transient energy were obtained. All analytical predictions were verified numerically by simulating the equation of motion of the HCVI-oscillator. The results also point out the advantage of the energy-based methods for the prediction of response regimes in strongly non-linear dynamical systems, and action-angle variables in particular.

The aforementioned results give a full perspective about the vibration mitigation performances of an HCVI-based passive energy absorber (HCVI-NES) and can be used for accurate optimization of the absorber design parameters, prediction of its absorption rate for a given set of excitation properties, and hence, the resistance of the primary structure. In comparison to the cubic NES, the HCVI-NES can utilize the saddle mechanism as a feature to gaining substantially higher energy absorption capabilities and thus enhanced performances. In other terms, as a hybrid NES, the HCVI-NES uses the built-in property of adaptive nonlinearity: the coexistence of multiple amplitude-based nonlinearities that are "switched-on" when the energy of the system exceeds a chosen threshold and enables a broader efficiency range. In the current work, the HCVI-NES was considered, which is a hybridization between a cubic NES and a VI-NES. This NES design utilizes two nonlinearities, that stem from both the cubic nonlinearity and elastic collisions with rigid barriers. Both are hardening nonlinearities that lead to an increase in the response frequency. Further works will focus on the optimization of the hybrid NES and examine the contribution of a softening nonlinearity to the overall absorption capabilities of the nonlinear passive energy absorber.

We learn that the saddle mechanism can lead to catastrophic consequences in HCVI-like engineering systems, such as pre-tensioned cables, springs, and compliant composite materials under rigid mechanical constraints, that might potentially result in a drastic increase in the response energy due to a slight uncontrolled change in the forcing parameters and finally lead to mechanical wear and failure. On the other hand, for HCVI-NES the saddle mechanism allows significant enhancement in the absorption rate and thus improves vibration protection capabilities. Those findings and insights are to be considered in the design and optimization process of such a vibration absorber.

**Declarations**

**Funding** M. Farid has been supported by the Fulbright Program, the Israel Scholarship Education Foundation (ISEF), Jean De Gunzburg International Fellowship, the Israel Academy of Sciences and Humanities, the Yitzhak Shamir Postdoctoral Scholarship of the Israeli Ministry of Science and Technology, the PMRI – Peter Munk Research Institute - Technion, and the Israel Science Foundation Grant 1696/17.

**Availability of data and material** The data that supports the findings of this study is available from the author upon request.

## Appendix A. Detailed derivation of action-angle transformation

In this appendix, a detailed derivation of the action-angle transformation is described. Two distinct cases are considered. The first case corresponds to SNO regime and $E \in (0, 1)$. The second case corresponds to the HCVI-regime and $E > 1$. For the first case, the following expression is obtained:

$$I(E; E \in (0,1)) = \frac{4}{2\pi} \int_0^{\sqrt[4]{E}} 2(E - q^4) dq = \frac{4}{\sqrt{2\pi}} E^{\frac{3}{4}} \int_0^1 \sqrt{1 - u^4} du = \frac{4}{3\pi} \mathbf{K}\left(\frac{1}{\sqrt{2}}\right) E^{\frac{3}{4}} \quad (A.1)$$

Here $u = x/\sqrt[4]{E}$. For $E > 1$, the limits of the integral change and the following form is obtained:

$$I(E; E > 1) = \frac{4}{2\pi} \int_0^1 2(E - q^4) dq = \frac{4}{\sqrt{2\pi}} E^{\frac{3}{4}} \int_0^{1/\sqrt[4]{E}} \sqrt{1 - u^4} du = \frac{2\sqrt{2}}{\pi} {}_2\mathbf{F_1}\left(-\frac{1}{2}, \frac{1}{4}; \frac{5}{4}; \frac{1}{E}\right) \sqrt{E} \quad (A.2)$$

finally, the averaged action variable is defined in a piece-wise fashion, as shown in Eq. (10). Now, the angle variable is derived according to Eq. (4):

$$\theta = \frac{\partial}{\partial I} \int_0^q \sqrt{2(E - U(x))} dx = \omega(E) \frac{\partial}{\partial E} \int_0^q \sqrt{2(E - U(x))} dx = \frac{\omega(E)}{\sqrt{2E}} \int_0^q \frac{1}{\sqrt{1 - \frac{U(x)}{E}}} dx \quad (A.3)$$

Using Eq. (A.3), the angle variable is obtained as follows:

$$\theta = \frac{\omega(E)}{\sqrt{2}} E^{-\frac{1}{4}} \int_0^{q/\sqrt[4]{E}} \frac{1}{\sqrt{1 - u^4}} du = \frac{\omega(E)}{\sqrt{2}} E^{-\frac{1}{4}} \mathbf{F}(u, i)\bigg|_{u=0}^{q/\sqrt[4]{E}} = \frac{\omega(E)}{\sqrt{2}} E^{-\frac{1}{4}} \mathbf{F}(q/\sqrt[4]{E}, i) \quad (A.4)$$

Here $u = q/\sqrt[4]{E}$, $i = \sqrt{-1}$, and $\mathbf{F}$ is the incomplete elliptic integral of the first kind. By inverting Eq. (A.4) we obtain the following expression:

$$u(E, \theta) = \mathrm{sn}\left(\frac{\sqrt{2}\theta E^{\frac{1}{4}}}{\omega(E)}, i\right) \to q(E, \theta) = E^{\frac{1}{4}} \mathrm{sn}\left(\frac{\sqrt{2}\theta E^{\frac{1}{4}}}{\omega(E)}, i\right) \quad (A.5)$$

Fourier coefficients of the displacement $q(E, \theta)$ are retrieved from well-known nomal expansion of the elliptic function:

$$\mathrm{sn}(z, k) = \frac{2\pi}{\mathbf{K}(k)k} \sum_{n=0}^{\infty} \frac{Q^{n+1/2} \sin((2n+1)\zeta)}{1 - Q^{2n+1}}, \quad Q = \exp\left(\frac{-\pi \mathbf{K}(k')}{\mathbf{K}(k)}\right), \zeta = \frac{\pi z}{2\mathbf{K}(k)} \quad (A.6)$$

Here $k' = \sqrt{1 - k^2}$. Hence, the first term of the series is as follows:

$$\mathrm{sn}(z, k) = \frac{2\pi}{\mathbf{K}(k)k} \frac{\sqrt{Q(k)}}{1 - Q(k)} \sin(\zeta) \quad (A.7)$$

The first order approximation of the displacement can be written as follows:

$$q(E,\theta) = \mu(E)\sin(\gamma(E)\theta) = \sum_{n=1}^{\infty} a_n \sin(n\theta), \quad \mu(E) = \frac{2\pi\eta}{\mathbf{K}(i)(1+\eta^2)}E^{\frac{1}{4}} = \alpha E^{\frac{1}{4}} \quad \text{(A.8)}$$

Here $\eta = \sqrt{|Q(i)|} = e^{-\pi/2}$. The coefficient of the first term in the series is obtained as follows:

$$a_1(E) = \frac{1}{\pi}\int_{-\pi}^{\pi} \mu(E)\sin(\gamma(E)\theta)\sin(\theta)d\theta = \begin{cases} \mu(E) & , E \in [0,1) \\ \frac{2\mu(E)\sin(\pi\gamma(E))}{\pi(1-\gamma(E)^2)} & , E \in [1,\infty) \end{cases} \quad \text{(A.9)}$$

Here $\gamma(E)$ corresponds to Eq. (12).

# List of Abbreviations

| | |
|---|---|
| AA | action-angle (variables) |
| HCVI | Hybrid cubic vibro-impact (regime, oscillator, NES) |
| LPT | Limiting phase trajectory |
| MM | Maximum mechanism |
| MM0, MM$\pi$ | Maximum mechanism at $\nu = 0, \pi$, respectively |
| NES | nonlinear energy sink |
| PEA | Passive energy absorber |
| RM | Resonance manifold |
| SM | Saddle mechanism |
| SNO | Smooth nonlinear oscillations |
| TMD | Tuned mass damper |
| TRC | Transient resonance capture |

# List of Symbols

| | |
|---|---|
| $2d$ | The channel's length |
| $\bar{\omega}$ | Characteristic dimensional frequency parameter |
| $\bar{F}(t)$ | Dimensional time-dependant forcing |
| $\delta$ | The Dirac delta function |
| $(\dot{\cdot}), (\cdot)'$ | Differentiation with respect to time scale $\tau$ and to averaged energy $\xi$, respectively |
| $\kappa$ | Restitution coefficient |
| $\mathbf{K}, \mathbf{F}$ | The elliptic integral of the first kind and the incomplete elliptic integral of the first kind, respectively |
| $_{p_1}\mathbf{F}_{p_2}$ | The generalized hypergeometric function |
| $\nu$ | Phase variable |
| $\omega$ | Response frequency of the oscillator |
| $\Omega^*, f^*$ | Frequency and critical forcing amplitude associated with coexistence of both 'saddle mechanism' and 'maximum mechanism' |
| $\text{sn}(z, k)$ | The Jacobi elliptic function of module $k$ |
| $\tilde{\xi}$ | The maximal transient energy level |
| $\xi_s$ | The transient energy level associates with the saddle point of the RM |
| $a_k$ | The coefficient of the $k^{th}$ term in Fourier series of solution $q(I, \theta)$ |
| $C(\nu, \xi)$ | The expression describes the resonance manifold |
| $E, \xi$ | Instantaneous and averaged energy of the oscillator |
| $F, \bar{\Omega}$ | Dimensional forcing amplitude and frequency |
| $f, \Omega$ | Non-dimensional forcing amplitude and frequency |
| $f_m, f_s$ | The critical forcing amplitudes associated with the occurrence of bifurcation through through maximum and saddle mechanisms, respectively |
| $H$ | Integral of motion/conservation law |
| $H_0$ | Initial conditions-related value of the integral of motion $H$ |
| $i$ | Unit imaginary number |
| $I, \theta$ | Action and angle variables |
| $J$ | Averaged action variable |
| $k$ | Dimensional stiffness coefficient of the cubic spring |
| $m$ | Mass of the oscillator |
| $p$ | Momentum of the oscillator |
| $t, \tau$ | Dimensional and non-dimensional time scales |
| $U(q)$ | Equivalent potential energy function |
| $x, q$ | Dimensional and non-dimensional displacement of the oscillator |